\documentclass[review,12pt]{elsarticle}

\usepackage{caption,geometry}
\usepackage[colorlinks]{hyperref}

\journal{The Innovation}

\biboptions{numbers,sort&compress}
\newcommand{\upcitep}[1]{\textsuperscript{\textsuperscript{\cite{#1}}}}
\setcitestyle{open={},close={}}

\begin{document}

\begin{frontmatter}

\title{Evolution of the Toroidal Flux of CME Flux Ropes during Eruption}

\author{Chen Xing,$^{1,2}$ Xin Cheng,$^{1,2,*}$ and Mingde Ding$^{1,2}$}
\address{$^1$School of Astronomy and Space Science, Nanjing University, Nanjing, 210046, China}
\address{$^2$Key Laboratory of Modern Astronomy and Astrophysics (Nanjing University), Ministry of Education, Nanjing 210093, China}

\cortext[]{Corresponding author\\
\hspace*{0.5cm} \textit{Email Address:} \href{mailto:xincheng@nju.edu.cn}{xincheng@nju.edu.cn} (Xin Cheng)}

\begin{abstract}
Coronal mass ejections (CMEs) are large-scale explosions of the coronal magnetic field. It is believed that magnetic reconnection significantly builds up the core structure of CMEs, a magnetic flux rope, during the eruption. However, the quantitative evolution of the flux rope, particularly its toroidal flux, is still unclear. In this paper, we study the evolution of the toroidal flux of the CME flux rope for four events. The toroidal flux is estimated as the magnetic flux in the footpoint region of the flux rope, which is identified by a method that simultaneously takes the coronal dimming and the hook of the flare ribbon into account. We find that the toroidal flux of the CME flux rope for all four events shows a two-phase evolution: a rapid increasing phase followed by a decreasing phase. We further compare the evolution of the toroidal flux with that of the Geostationary Operational Environmental Satellites soft X-ray flux and find that they are basically synchronous in time, except that the peak of the former is somewhat delayed. The results suggest that the toroidal flux of the CME flux rope may be first quickly built up by the reconnection mainly taking place in the sheared overlying field and then reduced by the reconnection among the twisted field lines within the flux rope, as enlightened by a recent 3D magnetohydrodynamic simulation of CMEs.
\end{abstract}


\end{frontmatter}

\section{Introduction} \label{sec1}
Coronal mass ejections (CMEs) are the largest-scale eruptive processes in the solar system. After eruption, CMEs rapidly accelerate to a velocity of hundreds of kilometres per second, sometimes even up to 3000 km s$^{-1}$.\upcitep{Yashiro2004} Once arriving at Earth and interacting with the magnetosphere, they can give rise to severe impacts on the high-tech activities of humans.\upcitep{Gosling1993} It is generally believed that the key magnetic configuration of CMEs is a flux rope in which the magnetic field lines wrap around a central axis, as indicated by helical structures often observed in high-temperature extreme UV (EUV) and white-light images.\upcitep{Chen1997, Cheng2014, Vourlidas2013} It is even argued that the flux rope may exist prior to the eruption, as shown by filaments, hot channels, sigmoids, dark cavities and so on (see the review by Cheng et al.\upcitep{Cheng2017}).

Regardless of whether there is a pre-existing (or seed) flux rope, the CME flux rope will be further built up by magnetic reconnection during the eruption. The evolution of the CME flux rope is first qualitatively described by the 2D standard CME/flare model.\upcitep{Carmichael1964, Sturrock1966, Hirayama1974, Kopp1976, Shibata1999, Lin2000} According to the model, the erupting flux rope stretches the overlying field and induces flare reconnection underneath. The reconnection produces closed field lines that are added to the erupting flux rope above and post-flare loops below the reconnection site. It is obvious that such a 2D reconnection contributes to only the poloidal flux of the flux rope, and the toroidal flux of the CME flux rope is completely inherited from its pre-existing value.\upcitep{Lin2004, Qiu2007} However, in a 3D case, the overlying field is always sheared, and the reconnection should produce a toroidal component of the magnetic flux, as simulated in many 3D models.\upcitep{Manchester2004, Aulanier2010, Aulanier2012} Therefore, the toroidal flux of the flux rope is expected to increase during the eruption. Interestingly, in a recent 3D magnetohydrodynamic (MHD) simulation by Aulanier and Dudík,\upcitep{Aulanier2019} three magnetic reconnection processes were revealed during the eruption. In addition to the reconnection in the sheared overlying field ($aa$--$rf\ reconnection$), they found two new reconnection processes: the reconnection in the flux rope that produces a more twisted flux rope and post-flare loops ($rr$--$rf\ reconnection$), and the reconnection between the flux rope and ambient sheared arcades that forms a new flux rope and post-flare loops ($ar$--$rf\ reconnection$). It is expected that the toroidal flux of the flux rope will be reduced by the $rr$--$rf\ reconnection$ and maintained by the $ar$--$rf\ reconnection$, respectively.\upcitep{Aulanier2019} In this paper, we observationally quantify the evolution of the toroidal flux of the CME flux rope during the eruption. The results can be used to differentiate and test the models discussed above.

The toroidal flux of a CME flux rope can be estimated by measuring the magnetic flux in its footpoint regions. To identify the footpoint regions of the flux rope, two methods have been proposed, which take advantage of J-shaped flare ribbons and twin dimmings caused by the eruption, respectively. For the former, the footpoint regions are thought to correspond to the regions surrounded by the hooks of flare ribbons.\upcitep{Wang2017, Aulanier2019} The reason is that the footpoint regions are believed to be enclosed by the footprints of quasi-separatrix layers,\upcitep{Demoulin1996b, Titov2007, Pariat2012, Janvier2013} which are highly consistent with the J-shaped flare ribbons.\upcitep{Savcheva2015, Savcheva2016, Zhao2016} This finding was further confirmed by Cheng and Ding,\upcitep{ChengX2016} who found that the two footpoints of the CME flux rope that appears as an EUV hot channel are even co-spatial with the regions mostly surrounded by the two hooks of flare ribbons. For the latter, the eruption of CMEs leads to plasma rarefaction in the low corona that manifests as dimmings inside and/or around the source regions, as shown in the EUV passbands.\upcitep{Harra2001, Cheng2012, Tian2012} There are two types of coronal dimmings: twin (or core) dimmings, which are located at opposite magnetic polarities in the source region of the eruptions,\upcitep{Sterling1997, Webb2000} and remote (or secondary) dimmings, which are far away from the source region and are shallower than the former.\upcitep{Thompson2000, Mandrini2007, Dissauer2018} Twin dimmings are often believed to map the footpoints of the erupting flux rope.\upcitep{Sterling1997, Webb2000, ChengJX2016, Qiu2017, Wang2019} In this paper, to identify the footpoint regions of the CME flux rope with a higher accuracy, we propose a new method that combines the previous two methods, which is presented in detail in Section \ref{sec2}. We show the results about the evolution of the toroidal flux during the eruption in Section \ref{sec3}, which is followed by a discussion in Section \ref{sec4}.

\section{Material and Method} \label{sec2}
\begin{figure}
\centering
\includegraphics[width=15cm]{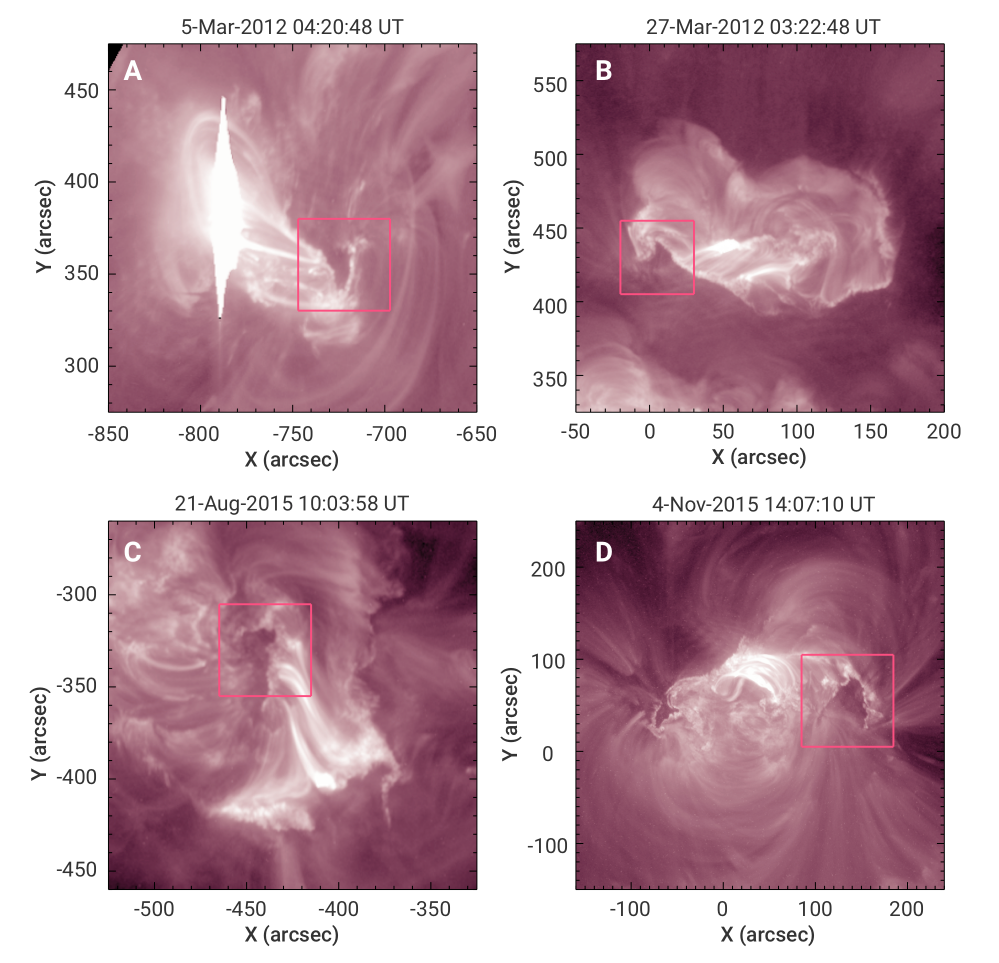}
\captionsetup{font=normalsize}
\caption{AIA 211 $\textup{\AA}$ images showing an overview of the four eruption events. The dark pink boxes indicate the field of view of the corresponding images in Figure \ref{fig4}. The box in (C) also indicates the field of view of each panel in Figure \ref{fig2}.}
\label{fig1}
\end{figure}

To quantify the toroidal flux of a CME flux rope, we measure the magnetic flux in its footpoint region. The footpoint region is defined as an isolated region that manifests as coronal dimming in EUV passbands and is surrounded by the flare ribbon hook. Here, we assume that the flux rope is not bifurcated. The events suitable for measurements should meet the following requirements. There is at least one J-shaped flare ribbon with an obvious hook. The region surrounded by the hook should present a clear dimming in the EUV passbands and not be obscured by the foreground, such as coronal loops and post-flare loops. In this study, the dimming region is detected at the 211 $\textup{\AA}$ passband, as widely used in previous studies.\upcitep{Zhang2017, Dissauer2018} In addition, the event should be located far away from the solar limb to ensure an accurate measurement of the magnetic field. Finally, we select four CME/flare events that meet the above requirements for this study. For cases 1 and 4 (2 and 3), we focus on only their western (eastern) footpoints. We do not consider their conjugate footpoints for the following reasons. The eastern footpoint for case 1 and the western footpoint for case 2 are obscured by post-flare loops. The dimming associated with the western footpoint for case 3 cannot be clearly identified, while the dimming within the eastern footpoint for case 4 seems to be mixed with the remote dimmings. The basic information of the four events is shown in Table \ref{tab1}. The 211 $\textup{\AA}$ images provided by the Atmospheric Imaging Assembly\upcitep{Lemen2012} (AIA) on board the Solar Dynamics Observatory\upcitep{Pesnell2012} ($SDO$), displaying the source regions of the four CMEs, are shown in Figure \ref{fig1}. The dark pink boxes mark the regions where the flux rope footpoints are located.

\begin{figure}
\centering
\includegraphics[width=15cm]{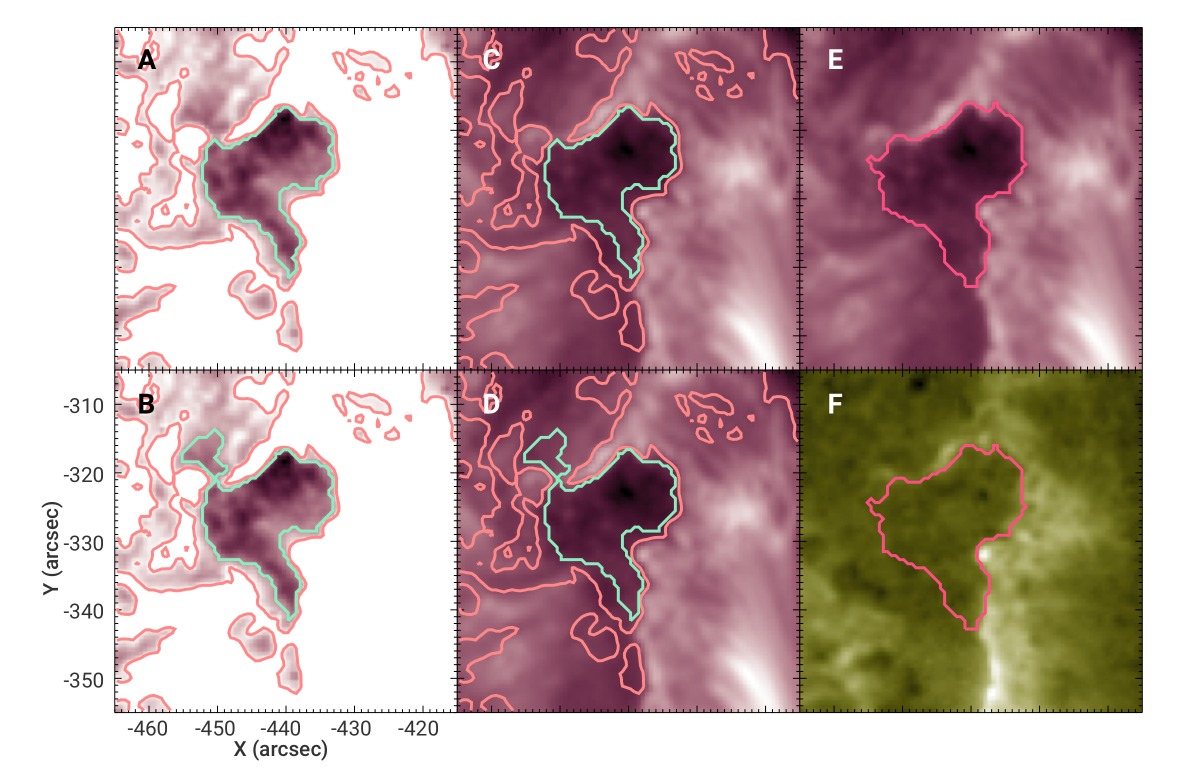}
\captionsetup{font=normalsize}
\caption{Example event (case 3) for interpreting the method of identifying the footpoint region of the CME flux rope. (A) The percentage difference image of the region in which the eastern footpoint is located at 09:55 UT, overlaid by the contours of $D=0$ (light pink) and the boundary of the grown region (green) with the threshold $T_{40}$. The region in white denotes the region of $D\ge0$. (B) The same as (A), but the boundary of the grown region is derived with the threshold $T_{39}$. (C) The same as (A), but the AIA 211 $\textup{\AA}$ image is used as the background. (D) The same as (B), but the AIA 211 $\textup{\AA}$ image is used as the background. The boundary of the identified footpoint region (dark pink) is overlaid on the AIA 211 $\textup{\AA}$ image (E) and the image of the AIA 1600 $\textup{\AA}$ to 1700 $\textup{\AA}$ intensity ratio (F). The boundary of the footpoint region adjacent to the ribbon hook matches well with the leading edge of the flare ribbon both in the 211 $\textup{\AA}$ passband and the 1600 $\textup{\AA}$ to 1700 $\textup{\AA}$ intensity ratio image.}
\label{fig2}
\end{figure}

In the following, we first introduce the method for identifying the footpoint region of the CME flux rope, taking case 3 that occurred on August 21, 2015 in National Oceanic and Atmospheric Administration (NOAA) Active Region 12403 as an example. Its footpoint region at 09:55 universal time (UT) is shown in Figure \ref{fig2} with the same field of view as that of the square in Figure \ref{fig1}C. To more clearly reveal the coronal dimming, we introduce the percentage difference ($D$) to show its relative change in intensity, which is defined by:
\begin{equation}
D_i=\frac{I_i-I_0}{I_0},
\end{equation}
where $I_i$ and $I_0$ are the intensities at 211 $\textup{\AA}$ at times during ($t_i$) and prior to ($t_0$) the eruption, respectively. At time $t_0$, no obvious brightening appears in the dimming region. Figure \ref{fig2}A shows the percentage difference image with contours of $D=0$ (light pink), which delineate the boundary of the dimming region. In addition, as the intensity at the flare ribbon hook increases but the intensity at the dimming region decreases during the eruption, the leading edge of the flare ribbon hook also refers to the boundary of the dimming region that the hook surrounds, as clearly shown in Figure \ref{fig2}C.

If the flare ribbon hook is closed, the dimming region surrounded by the hook naturally corresponds to the footpoint region of the flux rope.\upcitep{Wang2017} However, in most cases, the hook is only partially closed, as is shown in the example event. This phenomenon means that only part of the boundary of the footpoint region can be determined by the flare ribbon hook. For the remaining part of the boundary, we apply a region-grow method (performed by \texttt{REGION\_GROW} in IDL (Interactive Data Language)) to the percentage difference images. The method generates an isolated region (called the grown region) that starts from a seed and evolves until its boundary value reaches a pre-set threshold. With a given seed and a critical threshold, the method can generate a reference region that is as large as possible but still meets the following requirements: (1) the region is completely located at the concave side of the ribbon hook, (2) most (at least approximately half) of the boundary of the region matches the leading edge of the ribbon hook, and (3) the region does not contain subregions that are far away from the ribbon hook and connected to the main part by narrow corridors. Then, the reference region can help us replenish the remaining part of the boundary of the footpoint region.

To perform the above procedure, we need to determine the seed and the critical threshold. The first step is to approximately select a region that includes the footpoint region. The seed is then set to a subregion of 2$\times$2 pixels that contains the minimum value of $D$ (denoted by $D_s$) in the presupposed region and is obviously located at the concave side of the ribbon hook. To determine the critical threshold, we try 101 test thresholds ($T$):
\begin{equation}
T_j=\frac{j}{100}\times D_s\quad (j=0,1,2,...,100)
\end{equation}
Then, we search for the critical threshold $T_c$ and the corresponding reference region. The critical threshold is determined such that the grown region with $T_c$ meets the requirements for the reference region, while the grown region with $T_{c-1}$ extends dramatically and no longer meets these requirements. This method is practical since the relative change in the area of the grown region from $T_c$ to $T_{c-1}$ is larger than $10\%$ for a majority of the footpoint regions that we study.

The results derived by the region-grow method for case 3 are displayed in Figure \ref{fig2}A--\ref{fig2}D. In Figures \ref{fig2}A and \ref{fig2}B, we show the boundaries of the grown regions (green contours) generated with $T_{40}$ and $T_{39}$, respectively. They are also overlaid on the 211 $\textup{\AA}$ image, as shown in Figures \ref{fig2}C and \ref{fig2}D, respectively. Apparently, the grown region with $T_{40}$ meets the requirements for the reference region, while the grown region with $T_{39}$ has an obvious superfluous part that is not located at the concave side of the ribbon hook. Therefore, we consider $T_{40}$ as the critical threshold, which gives rise to a reasonable grown region, as shown in Figure \ref{fig2}C.

\begin{figure}
\centering
\includegraphics[width=15cm]{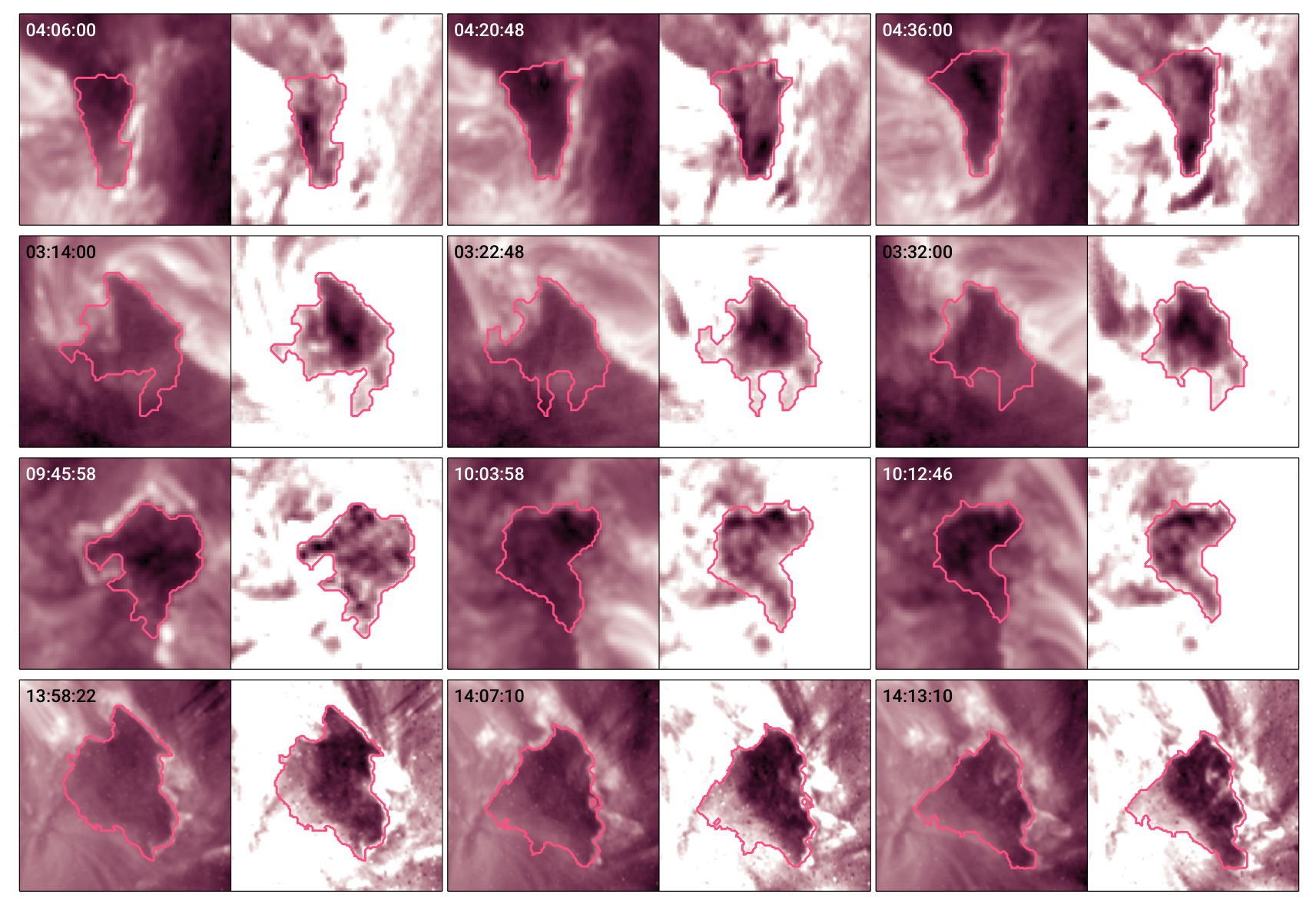}
\captionsetup{font=normalsize}
\caption{The boundaries of the identified footpoint regions of the CME flux ropes (dark pink) at three selected moments. The boundaries are overlaid on the AIA 211 $\textup{\AA}$ images in columns 1, 3, and 5, and overlaid on the percentage difference images in columns 2, 4, and 6, respectively. Rows 1--4 are for cases 1--4, respectively.}
\label{fig3}
\end{figure}

We can now determine the boundary of the footpoint region. On the side adjacent to the ribbon hook, the boundary of the footpoint region is mainly determined by the contour of $D=0$. For the remaining part of the boundary, it is determined by jointly considering the contour of $D=0$ and the boundary of the grown region. The boundary of the footpoint region for the example event is overlaid on the 211 $\textup{\AA}$ image (Figure \ref{fig2}E) and the image of the 1600 $\textup{\AA}$ to 1700 $\textup{\AA}$ intensity ratio (Figure \ref{fig2}F) that enhances the visibility of the flare ribbon.\upcitep{Dudik2016} It is worth mentioning that the contours of $D$ and the leading edge of the flare ribbon in the 211 $\textup{\AA}$ images detected by an edge-detection method (performed by \texttt{MORPH\_GRADIENT} in IDL) provide a double-check on the identification of the footpoint region. For each case in our study, we show the identified footpoint regions at three moments during the eruption (Figure \ref{fig3}). It is obvious that the footpoint regions map the main coronal dimming regions, and the majority of boundaries match well with the associated ribbon hooks.

Finally, the toroidal flux of the CME flux rope is estimated by integrating the radial component of the vector magnetic field ($B_r$) in the footpoint region. The vector magnetic field is obtained by the method in Guo et al.\upcitep{Guo2017} using the 720 s cadence data, which are observed by the Helioseismic and Magnetic Imager\upcitep{Scherrer2012} (HMI) on board the $SDO$ just before the flare start. To reduce the influence of noise in the magnetic field data, we integrate only the regions with $|B_{r}|>20$ G, following the threshold used by Wang et al.\upcitep{Wang2019} We multiply the projected area of each pixel by a factor of $1/\cos\theta$ to correct the projection effect of the area, where $\theta$ is the heliocentric angle of the pixel.

\section{Results} \label{sec3}
\begin{figure}
\centering
\includegraphics[width=15cm]{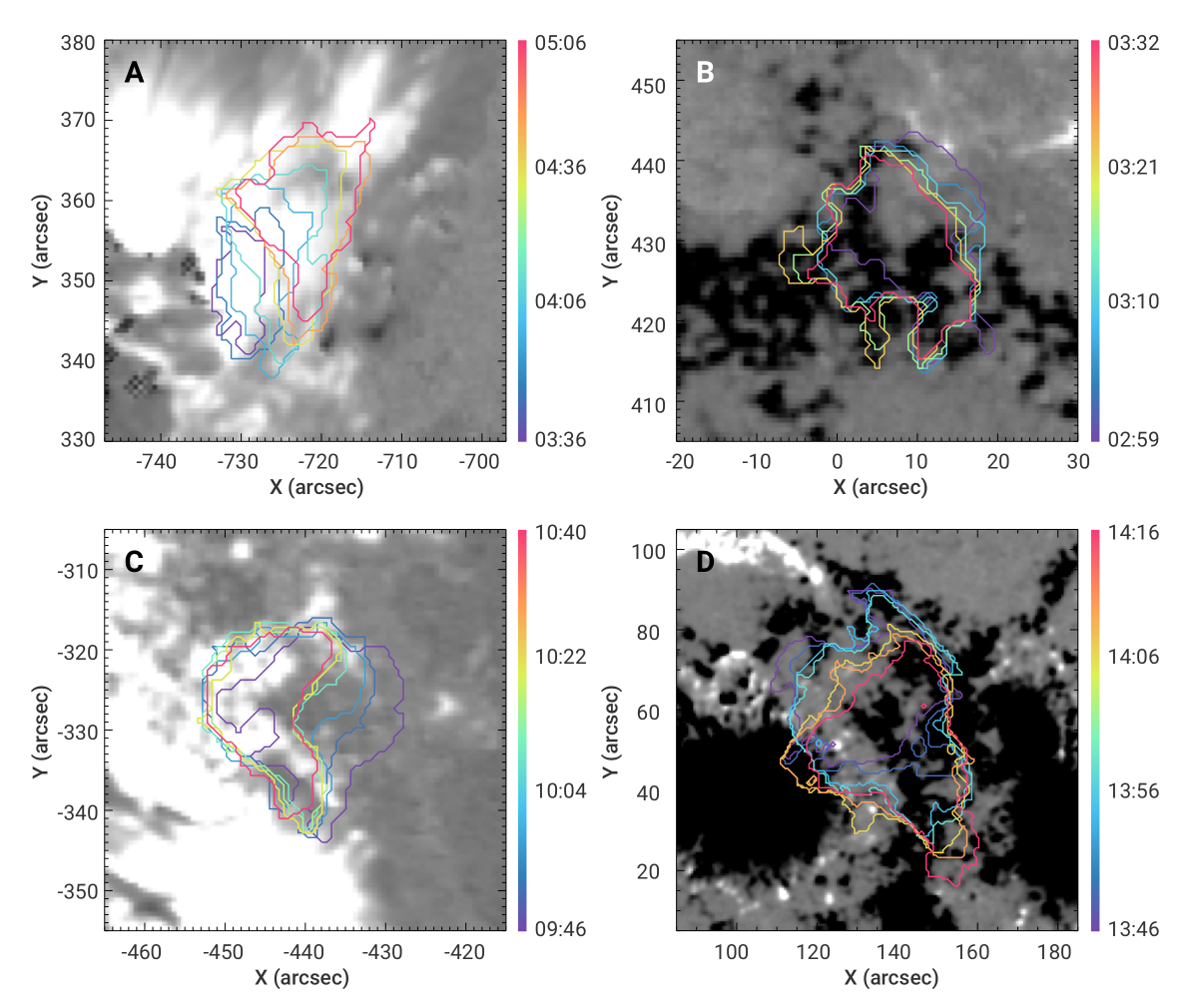}
\captionsetup{font=normalsize}
\caption{HMI radial magnetograms overlaid by the boundaries of the identified footpoint regions of the CME flux ropes. The evolution of the footpoint regions is denoted by different colours. (A)--(D) are for cases 1--4, respectively.}
\label{fig4}
\end{figure}

\begin{figure}
\centering
\includegraphics[width=15cm]{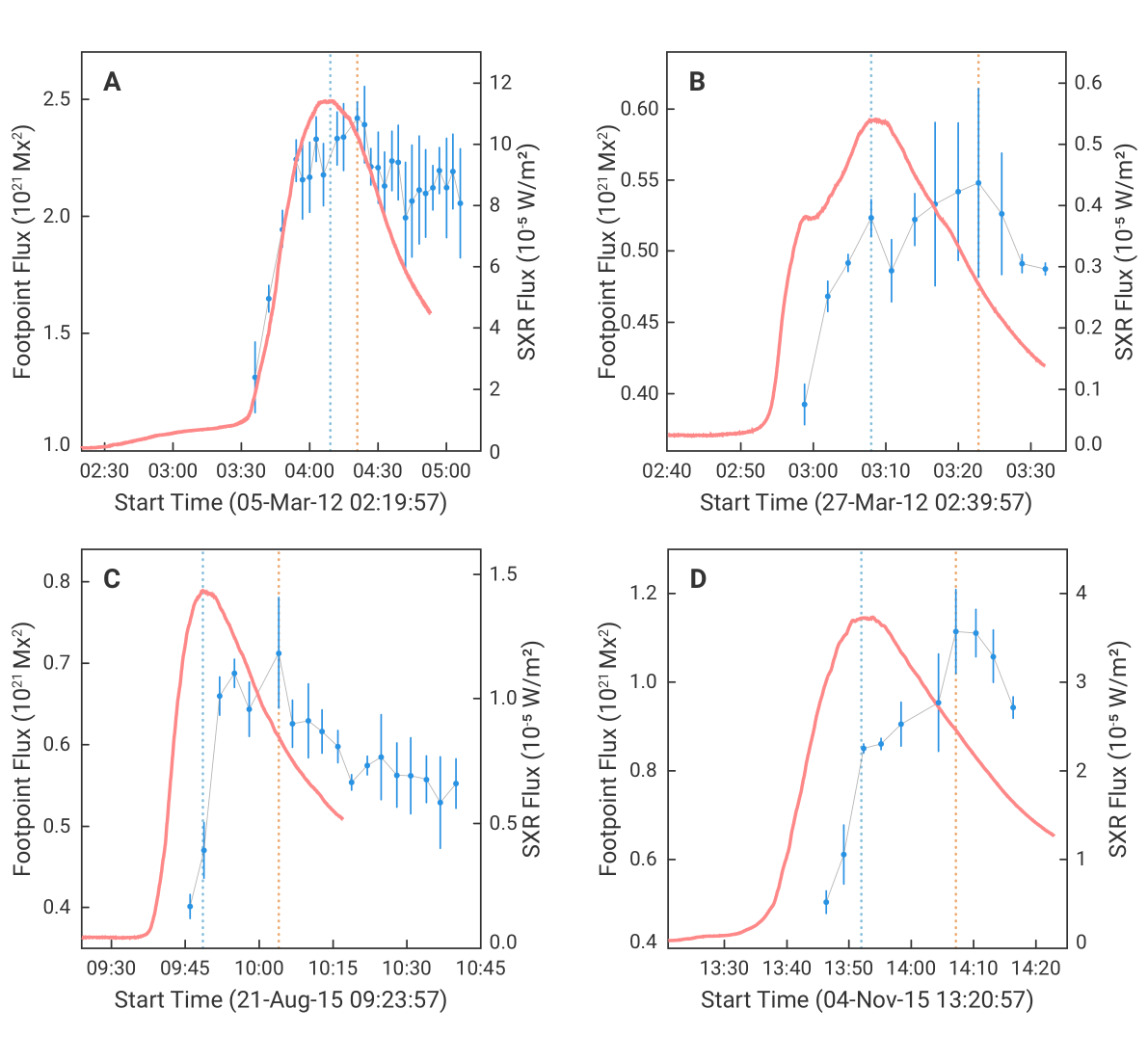}
\captionsetup{font=normalsize}
\caption{Temporal evolutions of the toroidal flux of the four CME flux ropes (blue) and the $GOES$ SXR 1--8 $\textup{\AA}$ flux of the associated flares (red). The error of the toroidal flux is from the standard deviation of multiple measurements. The orange and blue vertical dashed lines mark the peak times of the toroidal flux and the flare, respectively. (A)--(D) are for cases 1--4, respectively.}
\label{fig5}
\end{figure}

For each event, we measure the toroidal flux of the CME flux rope mainly at a time cadence of 3 minutes. The evolutions of the identified footpoint regions are demonstrated in Figure \ref{fig4}, which present obvious drifting of the flux rope footpoints as mentioned in previous papers.\upcitep{Aulanier2019, Chen2019, Lorincik2019, Zemanova2019} Figure \ref{fig5} shows the evolution of the toroidal flux during the eruption. It should be mentioned that our method sometimes derives some small and irregular dimming regions that extend from the main dimming region. It is difficult to judge whether these small regions belong to the footpoint region or not. Therefore, the boundary of the footpoint region is determined twice: once including these small regions and once without including them. For each moment, the final toroidal flux is the average of the measured fluxes in these two footpoint regions, and the corresponding error comes from the standard deviation. The maximum of the ratio of the error to the average flux is $\sim$12\%, indicating that the measured fluxes are not significantly influenced by manual factors. It is seen that for all the events in our study, the toroidal flux of the CME flux rope experiences a two-phase evolution during the eruption: a quick increasing phase followed by a decreasing phase, although the reduction in the toroidal flux for cases 1 and 2 seems to be less obvious than that in cases 3 and 4 considering that the absolute decrease is comparable to the errors. Quantitatively, for events 1, 2, 3, and 4, the toroidal flux increases by 84\%, 40\%, 78\%, and 121\%, respectively, relative to the first data point during the first phase; after the peak, it decreases by 15\%, 11\%, 23\%, and 15\%, respectively, relative to the peak flux during the second phase. Note that the period that we analyse covers only part of the CME evolution when the footpoint region can be clearly identified, and that the actual increase and decrease in the toroidal flux during the whole CME evolution could be larger than the deduced values. Interestingly, for case 4, the toroidal flux at 14:16 UT (the last moment in Figure \ref{fig5}D) is close to the toroidal flux of the associated magnetic cloud estimated by Wang et al.\upcitep{Wang2017}

An interesting finding is that the temporal evolution of the toroidal flux of the CME flux rope is roughly synchronous with that of the Geostationary Operational Environmental Satellites ($GOES$) soft X-ray (SXR) flux of the associated flare. Qualitatively, both of them show a quick increase followed by a decrease. Quantitatively, the increase in the toroidal flux mainly takes place during the flare rise phase, which amounts to 85\%, 84\%, and 57\% of the total increase for cases 1, 2, and 4, respectively. Considering that the CME flux rope may have evolved for a while before the first point of measurement, this ratio could be even larger. This finding indicates that the build-up of the CME toroidal flux is highly related to the rise phase of the associated flare when the energy is mostly released. Nevertheless, we also find a time delay in the peak of the toroidal flux relative to the peak of the $GOES$ SXR flux. After the flare peak, the SXR flux starts to decline, but the toroidal flux continues to increase for $\sim$10--20 minutes. It should be pointed out that, in our measurement, the toroidal flux of case 3 mainly increases after the flare peak. A possible reason is that our measurement for this case covers a shorter period of the rise phase. It is likely that the toroidal flux already has significantly increased before the first measurement point, and thus, the main increase in the toroidal flux still takes place during the rise phase, which is similar to the other three cases.

\section{Discussion} \label{sec4}
In this paper, we quantify the toroidal flux of a CME flux rope with a practical method of identifying the footpoint region of the flux rope. It is revealed that the toroidal flux first quickly increases and then decreases during the eruption for all four of the events we studied. The more important finding is that the evolution of the toroidal flux is generally synchronized with the evolution of the SXR flux, but the peak time of the former is $\sim$10--20 minutes later than that of the latter. The recent 3D MHD simulation of the CME performed by Aulanier and Dudík\upcitep{Aulanier2019} presents one of the models that may explain these results. In the context of the paper by Aulanier and Dudík,\upcitep{Aulanier2019} the increase and decrease in the toroidal flux may be related to the reconnection in the sheared overlying field ($aa$--$rf\ reconnection$) and that in the flux rope field ($rr$--$rf\ reconnection$), respectively. As mentioned in the Introduction, the $aa$--$rf\ reconnection$ can definitely increase the toroidal flux of the flux rope by converting the flux of the sheared overlying field into that of the flux rope. The $rr$--$rf\ reconnection$, as shown by Aulanier and Dudík,\upcitep{Aulanier2019} takes place between two field lines of the flux rope that are anchored close to the boundary of the footpoint region, producing a multi-turn flux rope field line and a post-flare loop. In this process, part of the flux of the flux rope can be transferred to the post-flare loop, thus resulting in a decrease in the toroidal flux. This process may account for the observational phenomenon that the flare ribbon sweeps and erodes the footpoint of the flux rope, which could lead to a contraction of the footpoint region and thus a reduction in the corresponding magnetic flux, as revealed in our study. By further combining the evolution of the associated flare, we speculate that these CME/flare events may experience the following three processes during the eruption. First, in the flare rise phase, the reconnection mainly takes place underneath the erupting flux rope, which quickly converts the sheared overlying flux into the flux rope and increases its toroidal flux. At the same time, the reconnection effectively heats the plasma and enhances the flare emission. After the flare peak, a similar but less energetic reconnection process continuously occurs in the early stage of the gradual phase, further increasing the toroidal flux of the flux rope but failing to sustain the increase in the SXR flux. Many studies\upcitep{Woods2011, Dai2013} have revealed that a less energetic reconnection could occur after the flare peak (in the EUV late phase of flares) but does not give rise to a significant enhancement of the SXR flux. The above two processes may correspond to the increasing phase of the toroidal flux. Finally, in the later gradual phase, the eruption process may be dominated by an even weaker reconnection in the flux rope field, which leads to a gradual decrease in the toroidal flux of the CME flux rope. Note that the eruption of the CME flux rope may also involve the reconnection between the flux rope and the ambient sheared arcades that are anchored close to the footpoint, which, however, is believed to lead to only some drifting of the footpoint without a change in the toroidal flux of the flux rope.\upcitep{Aulanier2019} It should be mentioned that in some cases, a small part of the CME flux rope footpoint may jump from one place to another rather than drift gradually when the flux rope reconnects with the ambient field that is rooted far away from the footpoint.\upcitep{Zhong2019} This phenomenon might also lead to a reduction in the toroidal flux as derived here.

We note that the evolution of the toroidal flux of the CME flux rope during the eruption revealed here has been suggested by previous papers. Aiming at the drifting of the footpoints of the CME flux rope, Chen et al.\upcitep{Chen2019} used a trial-and-error method to detect the core-dimming region and found that the magnetic flux there, i.e., the toroidal flux, shows a decreasing trend, if the core-dimming region is considered the footpoint region of the CME flux rope. In addition, Wang et al.\upcitep{Wang2017} studied the evolution of the twist of a CME flux rope and estimated its toroidal flux during the eruption. Their result also shows that the toroidal flux first increases and then decreases (see Figure 6 in Wang et al.\upcitep{Wang2017}). In fact, case 4 in the current study is the same as the event studied by Wang et al.\upcitep{Wang2017} We notice some quantitative differences between the toroidal flux obtained in our study and that in Wang et al.,\upcitep{Wang2017} especially in the decreasing phase. This finding can mainly be ascribed to the different methods of identifying footpoints used in the two studies. In the later stage of the event, Wang et al.\upcitep{Wang2017} assumed that only the brightening segment of the boundary of the flux rope footpoint evolves with time, but the dimmed segment remains unchanged. By comparison, the whole boundary evolves over time in the framework of our method. In addition, these two studies detect coronal dimmings at different passbands, which may also lead to a slight difference in the derived toroidal flux.

The method of identifying the footpoint region of the CME flux rope, which is the cornerstone of our study, is different from the previous methods. First, our method is also applicable to events that include partially closed ribbon hooks, which is a significant improvement to the methods that can be applied only to events with closed hooks.\upcitep{Wang2017} Second, our method has more geometrical restrictions on the dimming region, which is often determined with empirical thresholds for the absolute intensity or the difference intensity.\upcitep{Qiu2007, Reinard2008, Attrill2010, Dissauer2018} However, it can more accurately identify the footpoint regions of the flux rope, which are co-spatial with the twin dimming regions rather than the remote ones. If the remote dimming regions cannot be clearly separated from the twin dimming regions, their mixture\upcitep{Dissauer2018} can lead to an inaccurate estimation of the toroidal flux of the CME flux rope and its evolution. Our method can thus avoid such cases and give relatively accurate results.

\section*{Acknowledgements}
We are very grateful to the three referees for their helpful suggestions, which improve our manuscript. $SDO$ is a mission of NASA’s Living With a Star Program. C.X., X.C. and M.D.D. are funded by NSFC grants 11722325, 11733003, 11790303, 11790300, Jiangsu NSF grant BK20170011, and ``Dengfeng B" program of Nanjing University.

\section*{Declaration of Interests}
The authors declare no competing interests.

\clearpage
\newpage

\clearpage
\newpage

\clearpage
\newpage

\clearpage
\newpage

\clearpage
\newpage

\clearpage
\newpage
\begin{table}
\captionsetup{font=normalsize}
\caption{Parameters of the four CME/flare events}
\centering
\begin{tabular}{cccccc}
\hline
\hline
Number& Magnitude& NOAA&  Date&      Start Time& Peak Time\\
\hline
Flare 1&   X1.1& 11429& 2012/3/5&  02:30&                04:09               \\
Flare 2&   C5.3& 11444& 2012/3/27& 02:50&                03:08               \\
Flare 3&   M1.4& 12403& 2015/8/21& 09:34&                09:48               \\
Flare 4&   M3.7& 12443& 2015/11/4& 13:31&                13:52               \\
\hline
\\
\end{tabular}
\label{tab1}
\end{table}


\begin{thebibliography}{99}
\bibitem[Yashiro et al.(2004)]{Yashiro2004} Yashiro, S., Gopalswamy, N., Michalek, G., et al.\ (2004). A catalog of white light coronal mass ejections observed by the SOHO spacecraft. Journal of Geophysical Research (Space Physics) 109, A07105.
\bibitem[Gosling(1993)]{Gosling1993} Gosling, J.~T.\ (1993). The solar flare myth. Journal of Geophysical Research 98, 18937-18950.
\bibitem[Chen et al.(1997)]{Chen1997} Chen, J., Howard, R.~A., Brueckner, G.~E., et al.\ (1997). Evidence of an Erupting Magnetic Flux Rope: LASCO Coronal Mass Ejection of 1997 April 13. The Astrophysical Journal 490, L191-L194.
\bibitem[Cheng et al.(2014)]{Cheng2014} Cheng, X., Ding, M.~D., Guo, Y., et al.\ (2014). Tracking the Evolution of a Coherent Magnetic Flux Rope Continuously from the Inner to the Outer Corona. The Astrophysical Journal 780, 28.
\bibitem[Vourlidas et al.(2013)]{Vourlidas2013} Vourlidas, A., Lynch, B.~J., Howard, R.~A., et al.\ (2013). How Many CMEs Have Flux Ropes? Deciphering the Signatures of Shocks, Flux Ropes, and Prominences in Coronagraph Observations of CMEs. Solar Physics 284, 179-201.
\bibitem[Cheng et al.(2017)]{Cheng2017} Cheng, X., Guo, Y., and Ding, M.\ (2017). Origin and Structures of Solar Eruptions I: Magnetic Flux Rope. Science China Earth Sciences 60, 1383-1407.
\bibitem[Carmichael(1964)]{Carmichael1964} Carmichael, H.\ (1964). A Process for Flares. NASA Special Publication 50, 451.
\bibitem[Sturrock(1966)]{Sturrock1966} Sturrock, P.~A.\ (1966). Model of the High-Energy Phase of Solar Flares. Nature 211, 695-697.
\bibitem[Hirayama(1974)]{Hirayama1974} Hirayama, T.\ (1974). Theoretical Model of Flares and Prominences. I: Evaporating Flare Model. Solar Physics 34, 323-338.
\bibitem[Kopp \& Pneuman(1976)]{Kopp1976} Kopp, R.~A. and Pneuman, G.~W.\ (1976). Magnetic reconnection in the corona and the loop prominence phenomenon.. Solar Physics 50, 85-98.
\bibitem[Shibata(1999)]{Shibata1999} Shibata, K.\ (1999). Evidence of Magnetic Reconnection in Solar Flares and a Unified Model of Flares. Astrophysics and Space Science 264, 129-144.
\bibitem[Lin \& Forbes(2000)]{Lin2000} Lin, J. and Forbes, T.~G.\ (2000). Effects of reconnection on the coronal mass ejection process. Journal of Geophysical Research 105, 2375-2392.
\bibitem[Lin et al.(2004)]{Lin2004} Lin, J., Raymond, J.~C., and van Ballegooijen, A.~A.\ (2004). The Role of Magnetic Reconnection in the Observable Features of Solar Eruptions. The Astrophysical Journal 602, 422-435.
\bibitem[Qiu et al.(2007)]{Qiu2007} Qiu, J., Hu, Q., Howard, T.~A., et al.\ (2007). On the Magnetic Flux Budget in Low-Corona Magnetic Reconnection and Interplanetary Coronal Mass Ejections. The Astrophysical Journal 659, 758-772.
\bibitem[Manchester et al.(2004)]{Manchester2004} Manchester, W., Gombosi, T., DeZeeuw, D., et al.\ (2004). Eruption of a Buoyantly Emerging Magnetic Flux Rope. The Astrophysical Journal 610, 588-596.
\bibitem[Aulanier et al.(2010)]{Aulanier2010} Aulanier, G., T{\"o}r{\"o}k, T., D{\'e}moulin, P., et al.\ (2010). Formation of Torus-Unstable Flux Ropes and Electric Currents in Erupting Sigmoids. The Astrophysical Journal 708, 314-333.
\bibitem[Aulanier et al.(2012)]{Aulanier2012} Aulanier, G., Janvier, M., and Schmieder, B.\ (2012). The standard flare model in three dimensions. I. Strong-to-weak shear transition in post-flare loops. Astronomy and Astrophysics 543, A110.
\bibitem[Aulanier \& Dud{\'\i}k(2019)]{Aulanier2019} Aulanier, G. and Dud{\'\i}k, J.\ (2019). Drifting of the line-tied footpoints of CME flux-ropes. Astronomy and Astrophysics 621, A72.
\bibitem[Wang et al.(2017)]{Wang2017} Wang, W., Liu, R., Wang, Y., et al.\ (2017). Buildup of a highly twisted magnetic flux rope during a solar eruption. Nature Communications 8, 1330.
\bibitem[D{\'e}moulin et al.(1996)]{Demoulin1996b} D{\'e}moulin, P., Priest, E.~R., and Lonie, D.~P.\ (1996). Three-dimensional magnetic reconnection without null points 2. Application to twisted flux tubes. Journal of Geophysical Research 101, 7631-7646.
\bibitem[Titov(2007)]{Titov2007} Titov, V.~S.\ (2007). Generalized Squashing Factors for Covariant Description of Magnetic Connectivity in the Solar Corona. The Astrophysical Journal 660, 863-873.
\bibitem[Pariat \& D{\'e}moulin(2012)]{Pariat2012} Pariat, E. and D{\'e}moulin, P.\ (2012). Estimation of the squashing degree within a three-dimensional domain. Astronomy and Astrophysics 541, A78.
\bibitem[Janvier et al.(2013)]{Janvier2013} Janvier, M., Aulanier, G., Pariat, E., et al.\ (2013). The standard flare model in three dimensions. III. Slip-running reconnection properties. Astronomy and Astrophysics 555, A77.
\bibitem[Savcheva et al.(2015)]{Savcheva2015} Savcheva, A., Pariat, E., McKillop, S., et al.\ (2015). The Relation between Solar Eruption Topologies and Observed Flare Features. I. Flare Ribbons. The Astrophysical Journal 810, 96.
\bibitem[Savcheva et al.(2016)]{Savcheva2016} Savcheva, A., Pariat, E., McKillop, S., et al.\ (2016). The Relation between Solar Eruption Topologies and Observed Flare Features. II. Dynamical Evolution. The Astrophysical Journal 817, 43.
\bibitem[Zhao et al.(2016)]{Zhao2016} Zhao, J., Gilchrist, S.~A., Aulanier, G., et al.\ (2016). Hooked Flare Ribbons and Flux-rope-related QSL Footprints. The Astrophysical Journal 823, 62.
\bibitem[Cheng \& Ding(2016)]{ChengX2016} Cheng, X. and Ding, M.~D.\ (2016). The Characteristics of the Footpoints of Solar Magnetic Flux Ropes during Eruptions. The Astrophysical Journal Supplement Series 225, 16.
\bibitem[Harra \& Sterling(2001)]{Harra2001} Harra, L.~K. and Sterling, A.~C.\ (2001). Material Outflows from Coronal Intensity ``Dimming Regions'' during Coronal Mass Ejection Onset. The Astrophysical Journal 561, L215-L218.
\bibitem[Cheng et al.(2012)]{Cheng2012} Cheng, X., Zhang, J., Saar, S.~H., et al.\ (2012). Differential Emission Measure Analysis of Multiple Structural Components of Coronal Mass Ejections in the Inner Corona. The Astrophysical Journal 761, 62.
\bibitem[Tian et al.(2012)]{Tian2012} Tian, H., McIntosh, S.~W., Xia, L., et al.\ (2012). What can We Learn about Solar Coronal Mass Ejections, Coronal Dimmings, and Extreme-ultraviolet Jets through Spectroscopic Observations?. The Astrophysical Journal 748, 106.
\bibitem[Sterling \& Hudson(1997)]{Sterling1997} Sterling, A.~C. and Hudson, H.~S.\ (1997). Yohkoh SXT Observations of X-Ray ``Dimming'' Associated with a Halo Coronal Mass Ejection. The Astrophysical Journal 491, L55-L58.
\bibitem[Webb et al.(2000)]{Webb2000} Webb, D.~F., Lepping, R.~P., Burlaga, L.~F., et al.\ (2000). The origin and development of the May 1997 magnetic cloud. Journal of Geophysical Research 105, 27251-27260.
\bibitem[Thompson et al.(2000)]{Thompson2000} Thompson, B.~J., Cliver, E.~W., Nitta, N., et al.\ (2000). Coronal dimmings and energetic CMEs in April-May 1998. Geophysical Research Letters 27, 1431-1434.
\bibitem[Mandrini et al.(2007)]{Mandrini2007} Mandrini, C.~H., Nakwacki, M.~S., Attrill, G., et al.\ (2007). Are CME-Related Dimmings Always a Simple Signature of Interplanetary Magnetic Cloud Footpoints?. Solar Physics 244, 25-43.
\bibitem[Dissauer et al.(2018)]{Dissauer2018} Dissauer, K., Veronig, A.~M., Temmer, M., et al.\ (2018). On the Detection of Coronal Dimmings and the Extraction of Their Characteristic Properties. The Astrophysical Journal 855, 137.
\bibitem[Cheng \& Qiu(2016)]{ChengJX2016} Cheng, J.~X. and Qiu, J.\ (2016). The Nature of CME-flare-Associated Coronal Dimming. The Astrophysical Journal 825, 37.
\bibitem[Qiu \& Cheng(2017)]{Qiu2017} Qiu, J. and Cheng, J.\ (2017). Gradual Solar Coronal Dimming and Evolution of Coronal Mass Ejection in the Early Phase. The Astrophysical Journal 838, L6.
\bibitem[Wang et al.(2019)]{Wang2019} Wang, W., Zhu, C., Qiu, J., et al.\ (2019). Evolution of a Magnetic Flux Rope toward Eruption. The Astrophysical Journal 871, 25.
\bibitem[Zhang et al.(2017)]{Zhang2017} Zhang, Q.~M., Su, Y.~N., and Ji, H.~S.\ (2017). Pre-flare coronal dimmings. Astronomy and Astrophysics 598, A3.
\bibitem[Lemen et al.(2012)]{Lemen2012} Lemen, J.~R., Title, A.~M., Akin, D.~J., et al.\ (2012). The Atmospheric Imaging Assembly (AIA) on the Solar Dynamics Observatory (SDO). Solar Physics 275, 17-40.
\bibitem[Pesnell et al.(2012)]{Pesnell2012} Pesnell, W.~D., Thompson, B.~J., and Chamberlin, P.~C.\ (2012). The Solar Dynamics Observatory (SDO). Solar Physics 275, 3-15.
\bibitem[Dud{\'\i}k et al.(2016)]{Dudik2016} Dud{\'\i}k, J., Polito, V., Janvier, M., et al.\ (2016). Slipping Magnetic Reconnection, Chromospheric Evaporation, Implosion, and Precursors in the 2014 September 10 X1.6-Class Solar Flare. The Astrophysical Journal 823, 41.
\bibitem[Guo et al.(2017)]{Guo2017} Guo, Y., Cheng, X., and Ding, M.\ (2017). Origin and structures of solar eruptions II: Magnetic modeling. Science China Earth Sciences 60, 1408-1439.
\bibitem[Scherrer et al.(2012)]{Scherrer2012} Scherrer, P.~H., Schou, J., Bush, R.~I., et al.\ (2012). The Helioseismic and Magnetic Imager (HMI) Investigation for the Solar Dynamics Observatory (SDO). Solar Physics 275, 207-227.
\bibitem[Chen et al.(2019)]{Chen2019} Chen, H., Yang, J., Ji, K., et al.\ (2019). Observational Analysis on the Early Evolution of a CME Flux Rope: Preflare Reconnection and Flux Rope{\textquoteright}s Footpoint Drift. The Astrophysical Journal 887, 118.
\bibitem[L{\"o}rin{\v{c}}{\'\i}k et al.(2019)]{Lorincik2019} L{\"o}rin{\v{c}}{\'\i}k, J., Dud{\'\i}k, J., and Aulanier, G.\ (2019). Manifestations of Three-dimensional Magnetic Reconnection in an Eruption of a Quiescent Filament: Filament Strands Turning to Flare Loops. The Astrophysical Journal 885, 83.
\bibitem[Zemanov{\'a} et al.(2019)]{Zemanova2019} Zemanov{\'a}, A., Dud{\'\i}k, J., Aulanier, G., et al.\ (2019). Observations of a Footpoint Drift of an Erupting Flux Rope. The Astrophysical Journal 883, 96.
\bibitem[Woods et al.(2011)]{Woods2011} Woods, T.~N., Hock, R., Eparvier, F., et al.\ (2011). New Solar Extreme-ultraviolet Irradiance Observations during Flares. The Astrophysical Journal 739, 59.
\bibitem[Dai et al.(2013)]{Dai2013} Dai, Y., Ding, M.~D., and Guo, Y.\ (2013). Production of the Extreme-ultraviolet Late Phase of an X Class Flare in a Three-stage Magnetic Reconnection Process. The Astrophysical Journal 773, L21.
\bibitem[Zhong et al.(2019)]{Zhong2019} Zhong, Z., Guo, Y., Ding, M.~D., et al.\ (2019). Transition from Circular-ribbon to Parallel-ribbon Flares Associated with a Bifurcated Magnetic Flux Rope. The Astrophysical Journal 871, 105.
\bibitem[Reinard \& Biesecker(2008)]{Reinard2008} Reinard, A.~A., and Biesecker, D.~A.\ (2008). Coronal Mass Ejection-Associated Coronal Dimmings. The Astrophysical Journal 674, 576-585.
\bibitem[Attrill \& Wills-Davey(2010)]{Attrill2010} Attrill, G.~D.~R., and Wills-Davey, M.~J.\ (2010). Automatic Detection and Extraction of Coronal Dimmings from SDO/AIA Data. Solar Physics 262, 461-480.
\end{thebibliography}
\end{document}